\documentclass[preprint,authoryear,12pt]{elsarticle}

\usepackage{graphicx}

\usepackage{amssymb}

\usepackage[ps2pdf,%
a4paper=true,%
breaklinks=true,%
colorlinks=true,%
pdfauthor={G. La Mura et al.},%
pdftitle={Proceedings of the IX SCSLSA}%
]{hyperref}

\journal{Advances in Space Research}

\begin{document}

\newcommand{\ion}[2]{#1~{\small #2}}

\begin{frontmatter}



\title{{\bf The optical emission lines of type 1 X-ray bright Active Galactic Nuclei}}


\author{G. La Mura\corref{cor}}
\address{Institut f\"ur Astro- und Teilchenphysik -- Universit\"at Innsbruck, Technikerstr. 25/8, 6020 Innsbruck, Austria}
\cortext[cor]{Corresponding author}
\ead{giovanni.lamura@unipd.it}


\author{M. Berton, S. Ciroi, V. Cracco, F. Di Mille, P. Rafanelli\corref{}}
\address{Dipartimento di Fisica e Astronomia -- Universit\`a di Padova, Vicolo dell'Osservatorio 3, 35122 Padova, Italy}
\ead{marco.berton.1@studenti.unipd.it, stefano.ciroi@unipd.it, valentina.cracco@unipd.it, francesco.dimille@unipd.it, piero.rafanelli@unipd.it}


\begin{abstract}
A strong X-ray emission is one of the defining signatures of nuclear activity in galaxies. According to the Unified Model for Active Galactic Nuclei (AGN), both the X-ray radiation and the prominent broad emission lines, characterizing the optical and UV spectra of Type 1 AGNs, are originated in the innermost regions of the sources, close to the Super Massive Black Holes (SMBH), which power the central engine. Since the emission is concentrated in a very compact region (with typical size $r \leq 0.1\,$pc) and it is not possible to obtain resolved images of the source, spectroscopic studies of this radiation represent the only valuable key to constrain the physical properties of matter and its structure in the center of active galaxies. Based on previous studies on the physics of the Broad Line Region (BLR) and on the X-ray spectra of broad (FWHM$_{\rm H\beta} \geq 2000\,$km~s$^{-1}$) and narrow line (1000~km~s$^{-1} \leq$~FWHM$_{\rm H\beta} \leq 2000\,$km~s$^{-1}$) emitting objects, it has been observed that the kinematic and ionization properties of matter close to the SMBHs are related together, and, in particular, that ionization is higher in narrow line sources. Here we report on the study of the optical and X-ray spectra of a sample of Type 1 AGNs, selected from the Sloan Digital Sky Survey (SDSS) database, within an upper redshift limit of $z = 0.35$, and detected at X-ray energies. We present analysis of the broad emission line fluxes and profiles, as well as the properties of the X-ray continuum and Fe~K$\alpha$\ emission and we use these parameters to assess the consistency of our current AGN understanding.

\end{abstract}

\begin{keyword}
accretion, accretion disks; galaxies: active; galaxies: nuclei; quasars: emission lines; X-rays: galaxies
\end{keyword}

\end{frontmatter}

\parindent=0.5 cm

\section{Introduction}
It is nowadays well established that the nuclei of massive galaxies host Super Massive Black Holes (SMBH), with typical masses in the range of $10^6\, {\rm M_\odot} \leq M_{BH} \leq 10^9\, {\rm M_\odot}$, and that accretion of matter into the gravitational field of these objects leads to the triggering of nuclear activity. Active Galactic Nuclei (AGN, in brief) are known to irradiate the power of a whole galaxy ($L \geq 10^{44}\, {\rm erg\, s}^{-1}$) from a region of size smaller than 1~pc and to spread the spectral energy distribution (SED) of their output throughout the entire frequency range of the electro-magnetic radiation.

Although the high luminosity and the broad spectral distribution of the energy are quite general features of nuclear activity, there are various types of AGNs, with different spectroscopic and morphological characteristics. Based on the optical emission line profiles, \citet{Khachikian71,Khachikian74} proposed a classification scheme, further developed by \citet{Osterbrock81}, which essentially distinguished Seyfert galaxies in Type 1 sources, with broad permitted emission lines in their nuclear spectra, Type 2 objects, with only narrow emission lines, and a number of sub-classes having intermediate properties. Later on, the distinction between Type 1 and Type 2 objects was extended to other classes of AGNs and connected with the possibility to have a direct, unobscured view of the central energy source, from our observing position, in the framework of the well known AGN Unification models \citep[see e.g.][]{Antonucci85}.

In this contribution, we present the results of our ongoing investigation on the spectral properties of a sample of Type 1 AGNs, observed both in the optical and in X-rays, at energies of $0.2\, {\rm keV} \leq E \leq 12\, {\rm keV}$. Based on our previous results and on conclusions available in literature \citep[see e. g.][]{Rafanelli85, Popovic03, LaMura09a, LaMura09b}, we know that the optical emission line profiles and intensities are related to the structure and the physical conditions in the region where the emission of the broad spectral lines comes from (called Broad Line Region, or BLR). We show that the combined analysis of multiple frequency data provides a consistent picture of the physical processes, which are taking place close to the core of the active nucleus, where physics is very likely to be completely ruled by the SMBH and its accretion rate.

\begin{table}[t]
\caption{Properties of the sample objects}
\begin{footnotesize}
\begin{tabular}{lcccc}
\hline
Name & R.A. (J2000) & Dec. (J2000) & $z$ & ${F_X}^a$ \\
\hline
Mrk 1018 & 02:06:15.99 & $-$00:17:29.3 & 0.043 & $25.89 \pm 0.16$ \\
2MASX J03063958+0003426 & 03:06:39.58 & $+$00:03:43.2 & 0.107 & $1.73 \pm 0.18$ \\
2MASX J09043699+5536025 & 09:04:36.97 & $+$55:36:02.6 & 0.037 & $12.57 \pm 0.25$ \\
LEDA 26614 & 09:23:43.01 & $+$22:54:32.4 & 0.033 & $34.89 \pm 0.20$ \\
Mrk 110 & 09:25:12.87 & $+$52:17:10.5 & 0.035 & $59.31 \pm 0.15$ \\
NGC 3080 & 09:59:55.85 & $+$13:02:38.0 & 0.035 & $10.80 \pm 0.93$ \\
PG 1114+445 & 11:17:06.39 & $+$44:13:33.3 & 0.144 & $3.59 \pm 0.29$ \\
PG 1115+407 & 11:18:30.30 & $+$40:25:54.1 & 0.154 & $4.77 \pm 0.34$ \\
2E 1216.9+0700 & 12:19:30.87 & $+$06:43:34.8 & 0.080 & $1.86 \pm 0.04$ \\
Mrk 50 & 12:23:24.14 & $+$02:40:44.4 & 0.026 & $27.83 \pm 0.20$ \\
Was 61 & 12:42:10.61 & $+$33:17:02.6 & 0.043 & $13.11 \pm 0.29$ \\
LEDA 94626 & 13:48:34.99 & $+$26:31:09.3 & 0.059 & $2.94 \pm 0.02$ \\
PG 1352+183 & 13:54:35.69 & $+$18:05:17.5 & 0.151 & $5.60 \pm 0.05$ \\
Mrk 464 & 13:55:53.52 & $+$38:34:28.7 & 0.050 & $4.72 \pm 0.07$ \\
PG 1415+451 & 14:17:00.83 & $+$44:56:06.3 & 0.113 & $3.35 \pm 0.03$ \\
NGC 5548 & 14:17:59.51 & +25:08:12.4 & 0.016 & $74.44 \pm 0.15$ \\
NGC 5683 & 14:34:52.48 & $+$48:39:42.9 & 0.037 & $11.82 \pm 0.15$ \\
2MASS J14441467+0633067 & 14:44:14.67 & $+$06:33:06.7 & 0.208 & $5.43 \pm 0.05$ \\
Mrk 290 & 15:35:52.42 & $+$57:54:09.5 & 0.030 & $17.35 \pm 0.07$ \\
Mrk 493 & 15:59:09.67 & +35:01:47.3 & 0.031 & $14.02 \pm 0.07$ \\
2MASX J16174561+0603530 & 16:17:45.61 & $+$06:03:53.0 & 0.039 & $15.97 \pm 0.15$ \\
II Zw 177 & 22:19:18.53 & $+$12:07:53.2 & 0.081 & $4.66 \pm 0.04$ \\
Mrk 926 & 23:04:43.49 & $-$08:41:08.5 & 0.047 & $62.38 \pm 0.47$ \\
\hline
\end{tabular}
\\ $^a$ Fluxes expressed in units of $10^{-12}\, {\rm erg\, cm^{-2}\, s^{-1}}$
\end{footnotesize}
\end{table}
\section{Sample selection}
Due to the extremely peculiar phenomenology related to AGNs, investigation of nuclear activity is usually carried out either by focusing as much as possible all data gathering efforts on a few, well selected sources, or by considering very large samples of sources, whose properties can be, therefore, appreciably constrained by means of statistical considerations. In our work, we try to merge the two approaches, looking for a fair balance between spectral coverage and sample size. In particular, to investigate the central regions of AGNs, we require a sample of Type 1 objects, with optical broad emission line spectra, combined with the higher energy data that are more suitable to investigate the physics of the accretion process very close to the central engine.

In the optical range, the spectroscopic database of the Sloan Digital Sky Survey \citep[SDSS DR7, see][]{Abazajian09} represents a huge source of appreciable resolution optical spectra of point like targets ($1800 \leq R \leq 2000$). Combining this database with the scientific archive of the XMM Newton satellite, we have been able to identify a set of objects associated with a powerful X-ray emission. Since we are mainly interested on spectroscopic information, we restricted the X-ray selection band to the energy range of $0.2\, {\rm keV} \leq E \leq 12\, {\rm keV}$ and we only considered sources where we could clearly detect the Hydrogen Balmer emission line series, which has an important role in the determination of the BLR plasma physical conditions.

By searching the SDSS database for Type 1 AGN spectra with a signal to noise ratio larger than 10 in the continuum around 5100~\AA, in the redshift range $0.01 \leq z \leq 0.35$, where the Balmer series is still falling entirely in the optical domain, and with an associated X-ray flux of $F_X \geq 10^{-12}\, {\rm erg\, cm^{-2}\, s^{-1}}$, we identified 65 objects. Among these, 23, that we list in Table~1, had available X-ray spectra.

\section{Data reduction and measurements}
\subsection{Optical spectra}
The purpose of the optical spectra reduction was basically the extraction of pure BLR emission, trying to separate the Hydrogen Balmer lines from other spectral components. This task is not generally trivial, because the BLR signal is blended together with several other contributions to the observed light. Emission lines arising in the Narrow Line Region (NLR), which emits both permitted and forbidden lines, affect the observed profiles of the Balmer lines, while the continuum of the AGN and of its host galaxy stellar populations overlap with the broad emission lines, merging smoothly with their wings. The whole signal, finally, is prone to the effects of interstellar extinction, arising both in our own Galaxy, as well as from within the source itself.

\begin{figure}[t]
\includegraphics[width=\textwidth]{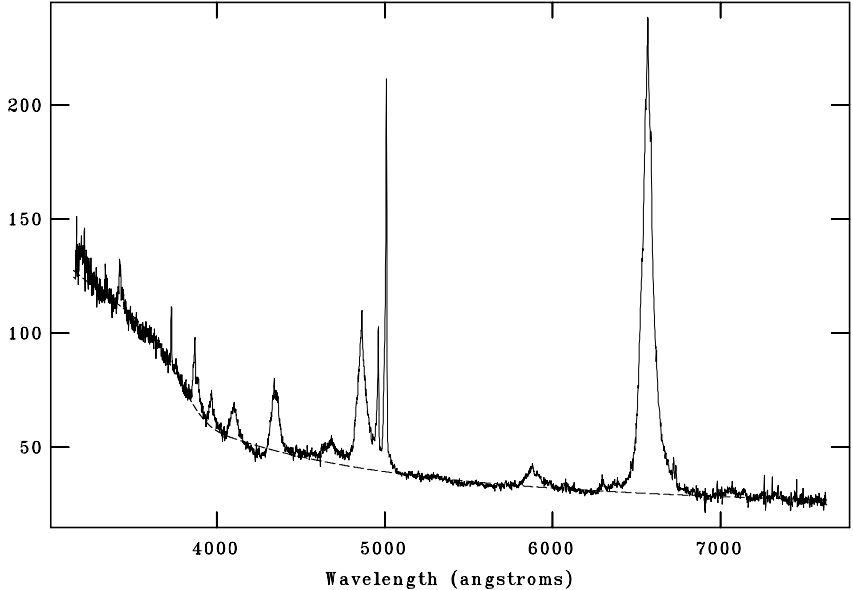}
\caption{Example of fit to the continuum of 2MASS J14441467+0633067 with a composite power-law, black body and Balmer continuum emission. Ordinates are specific fluxes in units of $10^{-17}\, {\rm erg\, cm^{-2}\, s^{-1}\, \AA^{-1}}$.}
\end{figure}
In order to extract as clean BLR spectra as possible, we used the \texttt{IRAF}\footnotemark \footnotetext{The \texttt{IRAF} project home page is available at \texttt{http://iraf.noao.edu/}} reduction software to deal with the SDSS data files. The first reduction step involved the correction for Galactic foreground extinction, which we evaluated by means of a parametric extinction curve \citep{Cardelli89}, calibrated with the $E(B-V)$ color excess computed on the SDSS object magnitudes and with the assumption of:
$$\frac{A_V}{E(B-V)} = 3.1 \eqno(1)$$
and applied with the task \texttt{deredden}. No considerations were made on intrinsic extinction, up to this point. As a second step, we took the spectra to their rest frame, by removing the effects of cosmological redshift, through the task \texttt{newredshift}. We finally re-scaled the spectra to a linear 1~\AA\ dispersion, with the task \texttt{dispcor}.

Once the spectra had been corrected for effects arising from outside the source, we moved to the estimation of the contributions overlapping with the BLR signal and of the emission line blends produced within the BLR itself. To estimate the underlying continuum, we applied a best fit procedure, combining a non-thermal power law, a black-body emission and a Balmer continuum emission, with the task \texttt{nfit1d}. An example of these fits is given in Fig.~1. The extraction of pure Balmer broad emission lines, on the other hand, was performed by means of multiple Gaussian profile fitting, applied with the task \texttt{ngaussfit}. In this case, we assumed the profiles of well isolated narrow emission lines, like [\ion{O}{III}] $\lambda\lambda 4959,5007$, to act as a template for the NLR spectral features, and we simply re-scaled its instances, in order to remove the emissions of [\ion{O}{III}] $\lambda 4363$, [\ion{N}{II}] $\lambda\lambda 6548,6584$, [\ion{O}{I}] $\lambda 6300$ and [\ion{S}{II}] $\lambda\lambda 6716,6731$. The fit to the profile of the Balmer lines was achieved with the combination of up to two Gaussian components, with the addition of further Gaussians in the case of severe blending of H$\beta$\ with \ion{He}{II} $\lambda 4686$. When strong \ion{Fe}{II} emissions were seen, we modeled their intensities following \citet{Kovacevic10}.

\subsection{X-ray spectra}
The X-ray spectra were obtained from the observation data packages provided by the XMM Newton Science Archive (XSA).\footnotemark \footnotetext{XSA is available at \texttt{http://xmm.esac.esa.int/xsa/}} The X-ray data reduction consists of the procedures we need to extract the spectra of the target source from the data tables produced by the instruments. These operations were carried out through the XMM Science Analysis System (\texttt{SAS} v. 10.0.0). At first, we selected the data tables produced by the EPIC-pn, MOS1 and MOS2 cameras, with the \texttt{odfbrowser} task. We, then, filtered the observations against high flaring particles backgrounds, identifying the source and background regions on the detectors within circular areas of 30'' of radius with \texttt{xmmselect}. We derived the instrument response and ancillary files for the observations with \texttt{rmfgen} and \texttt{arfgen} and, finally, we grouped the different channels, so that at least 20 counts for each data point where available, with \texttt{specgroup}.

\begin{figure}[t]
\includegraphics[angle=270, width=\textwidth]{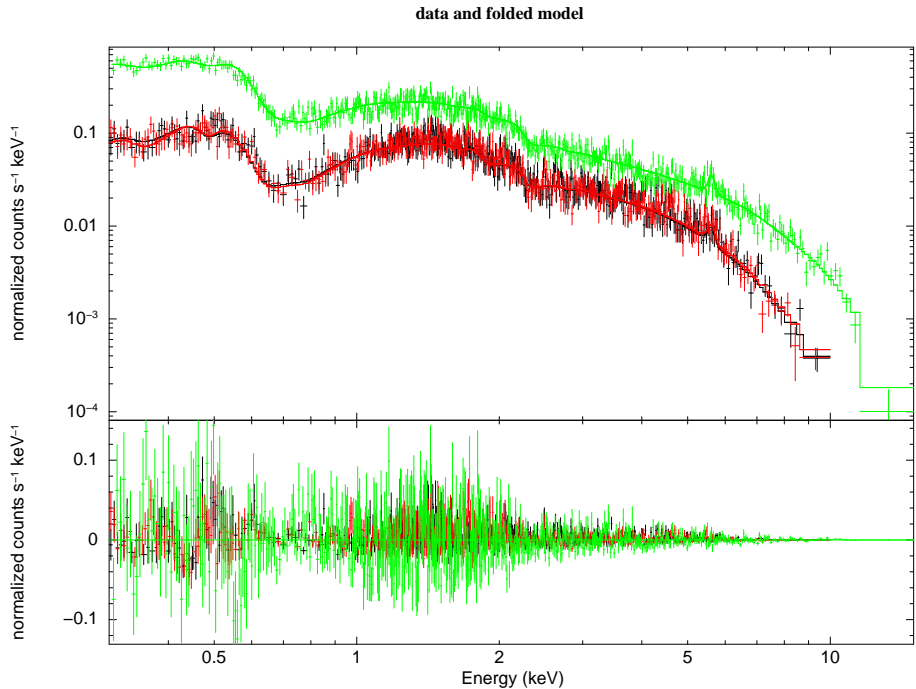}
\caption{{\bf Upper panel:} X-ray spectra of PG 1114+445 as observed with the EPIC-pn (upper spectrum) and by the twin MOS cameras (lower spectra) of XMM. The spectrum is modeled with a low energy thermal component, a power law emission, a Gaussian emission line and a noticeable ionized absorber. {\bf Lower panel:} Residuals of the model fit.}
\end{figure}
The extracted spectra, of which we give an example in Fig.~2, were modeled with the \texttt{xspec} software, through a combination of thermal and non-thermal components. The basic model had 4 mandatory components: i) a non-thermal power law emission; ii) a black-body thermal continuum; iii) an emission line with an initial rest frame energy of 6.4~keV, accounting for the Fe~K$\alpha$\ emission; iv) a neutral hydrogen absorption, matching the Galactic \ion{H}{I} column density estimated in the direction of the target \citep{Kalberla05}\footnotemark \footnotetext{Using the HEASARC service available at\\ \texttt{http://heasarc.gsfc.nasa.gov/cgi-bin/Tools/w3nh/w3nh.pl}}. The gas chemical composition is assumed solar. In case of unsatisfactory fits and, particularly, when the signal to noise ratio (s/n) was high enough to discriminate the existence of further components, we used additional contributions, such as composite thermal emission and neutral or ionized intrinsic absorptions, estimating the improvement of the models until we carried out fits to the observations with $\chi_{red}^2 \sim 1$ and model likelihood $P \geq 60\%$.

Assuming the redshift of the source and the neutral Hydrogen column density, corresponding to the line of sight within the Milky Way, to be fixed, the basic model had 7 free parameters to account for the temperature of the low energy thermal component, usually falling close to $k_B T \approx 0.1\,$keV, the photon index of the high energy power-law, ranging between $1 \leq \Gamma < 3$, and the rest frame energy and width of the Fe~K$\alpha$\ emission line, together with the normalization constants of these 3 contributions. In the case of detection of additional absorption contributions, generally suggested by difficulties to fit the low energy spectrum with the basic model, the column densities of neutral and ionized absorbers were also used as free parameters.

\subsection{Measurements}
Once the spectroscopic data had been reduced, we proceeded to the measurement of the Hydrogen Balmer emission line fluxes (up to H$\delta$\ or H$\epsilon$, when detected), to the estimate of FWHM$_{\rm H\beta}$\ and of the Fe~K$\alpha$\ rest frame energy.

Measurements on the optical spectra were collected, after evaluating the s/n in the adjacent continuum, by choosing 5 different levels of continuum, within reasonable ranges, and estimating the effects of such choices on the total flux and FWHM determinations. A convenient way to study the emission of Balmer lines from the BLR gas takes advantage from the Boltzmann Plot technique \citep[BP, see][]{Popovic03,Popovic06}. The method introduces a normalized line intensity for every transition $u \to l$ ($E_u > E_l$), through the relation:
$$I_n = \frac{F_{ul} \lambda_{ul}}{g_u A_{ul}}, \eqno(2)$$
where $\lambda_{ul}$ is the emission line wavelength, $A_{ul}$ its transition probability, $g_u$ the upper level's statistical weight and $F_{ul}$ the measured line flux. Since the fluxes express the energies carried away by radiation in the different emission lines, which are inversely proportional to the photon wavelengths, the normalized intensities are related to the number of photons emerging from the gas, due to the observed transitions, normalized with respect to the probability to emit one photon in that particular transition. The BP is the relationship between the normalized intensities of a series of transitions leading to the same lower level as a function of the upper level's excitation energy. It can be shown that an optically thin plasma, with a population of excited levels following the Boltzmann distribution, produces a BP of the form:
$$\log I_n = B - A \cdot E_u, \eqno(3)$$
where $E_u$ is the upper level's excitation energy and the parameter $A$ turns out to be inversely proportional to the plasma temperature. More generally, $A$ compares the strength of the emission lines in a series with the probability of emitting the corresponding photons. A large value of the parameter $A$ implies that the high order lines are considerably weaker than expected on the simple basis of their emission probability, while a low value expresses a relative enhancement of the high order lines. In the case of the broad component of the Balmer series, this parameter is related to the probability of destroying the transitions from high levels, through multiple photon scattering followed by de-excitations cascade. This effect is related to the number of neutral atoms in the medium and, therefore, to the ionization degree of the gas. Thus, lower values of $A$ suggest the presence of intense ionizing radiation fields, that can produce a large number of Balmer photons with a lower probability of re-absorption.

The X-ray Fe K$\alpha$\ emission, on the other hand, depends on the degree of ionization of heavy elements, which are the most important absorbers of very high frequency photons. The more penetrating hard X-ray radiation can travel into the deepest regions of the gas that are shielded from the lower energy ionizing photons and, therefore, host a predominantly neutral medium. Here, the hard X-ray radiation is able to produce Fe~K$\alpha$\ emission at energies which increase with the degree of ionization of Fe itself. If Fe is initially neutral, the process occurs at a rest frame energy of approximately 6.4~keV. Our measurements of the Fe K$\alpha$\ emission line energy relied on the model fit and the corresponding uncertainty calculations. The best fit value of the line centroid was assumed to be our highest confidence estimate, while variations of the model fit parameters within a 95\%\ confidence range yielded the uncertainty of the best fit parameter space.

\begin{table}[t]
\begin{center}
\caption{H$\beta$\ line profiles, Boltzmann Plot slopes and Fe emission line energy}
\begin{footnotesize}
\begin{tabular}{lccc}
\hline
Name & FWHM$_{\rm H\beta}$ & $A$ & $E_{K\alpha}$ \\
\hline
Mrk 1018 & 6940$\pm$760 & 0.44$\pm$0.11& 6.31$\pm$0.07\\
2MASX J03063958+0003426 & 2022$\pm$150 & 0.23$\pm$0.01& 6.59$\pm$0.18 \\
2MASX J09043699+5536025 & 6041$\pm$260 & 0.55$\pm$0.02 & 6.48$\pm$0.12 \\
LEDA 26614 & 2178$\pm$150 & 0.21$\pm$0.06 & 6.32$\pm$0.15 \\
Mrk 110 & 2316$\pm$158 & 0.64$\pm$0.02 & 6.45$\pm$0.03\\
NGC 3080 & 2015$\pm$150 & 0.09$\pm$0.05 & 6.57$\pm$0.09\\
PG 1114+445 & 6039$\pm$355 & 0.41$\pm$0.09 & 6.42$\pm$0.05 \\
PG 1115+407 & 2077$\pm$150 & 0.18$\pm$0.09 & 7.18$\pm$0.10 \\
2E 1216.9+0700 & 2043$\pm$150 & 0.23$\pm$0.06 & 6.48$\pm$0.40 \\
Mrk 50 & 5842$\pm$181 & 0.22$\pm$0.02 & 6.37$\pm$0.06 \\
Was 61 & 1776$\pm$150 & 0.62$\pm$0.01 & 6.53$\pm$0.10 \\
LEDA 94626 & 1820$\pm$150 & 0.35$\pm$0.01 & 6.52$\pm$0.16 \\
PG 1352+183 & 4527$\pm$290 & 0.21$\pm$0.09 & 6.45$\pm$0.24 \\
Mrk 464 & 5737$\pm$775 & 0.78$\pm$0.04 & 6.43$\pm$0.17 \\
PG 1415+451 & 3021$\pm$150 & 0.39$\pm$0.04 & 6.67$\pm$0.07 \\
NGC 5548 & 7228$\pm$409 & 0.64$\pm$0.10 & 6.40$\pm$0.03 \\
NGC 5683 & 4980$\pm$332 & 0.20$\pm$0.02 & 6.41$\pm$0.40 \\
2MASS J14441467+0633067 & 4285$\pm$305 & 0.40$\pm$0.02 & 6.42$\pm$0.20\\
Mrk 290 & 4586$\pm$177 & 0.29$\pm$0.06 & 6.37$\pm$0.06 \\
Mrk 493 & 1506$\pm$150 & 0.26$\pm$0.03 & 6.86$\pm$0.19 \\
2MASX J16174561+0603530 & 3283$\pm$527 & 0.47$\pm$0.05 & 6.50$\pm$0.11 \\
II Zw 177 & 1281$\pm$150 & 0.37$\pm$0.04 & 6.49$\pm$0.20 \\
Mrk 926 & 9035$\pm$250 & 0.74$\pm$0.11 & 6.43$\pm$0.11 \\
\hline
\end{tabular}
\end{footnotesize}
\end{center}
\end{table}
\section{Results and discussion}
A summary of determinations based on our Balmer emission line flux and profile measurements, together with the Fe~K$\alpha$\ emission line, is reported in Table~2. According to our interpretation of Type 1 AGN spectra, in which we assume the BLR to be mainly under the influence of the central engine, through the combined effects of its gravitational field and ionizing radiation,  we expect a relationship between the width of the broad line profiles and the central mass, in the form of:
$$M_{BH} = f \frac{R_{\rm BLR} \Delta v^2}{G}, \eqno(4)$$
where $f$ is a geometrical factor, typically estimated in the order of $\sim 5$ \citep{Onken04,LaMura09a}, in which we condensate the effects of the unknown structure of the BLR. The physical quantities appearing in Eq.~(4), however, cannot be directly estimated in single-epoch observations, and, even in the most advanced multi-epoch investigations, the actual definitions of the BLR radius and of its velocity field are still matter of debate. The problems which affect the determination of the mass in the central source reflect in the interpretation of its accretion rate and, therefore, on our expectations concerning the emitted spectrum.

\begin{figure}[t]
\includegraphics[width=0.48\textwidth]{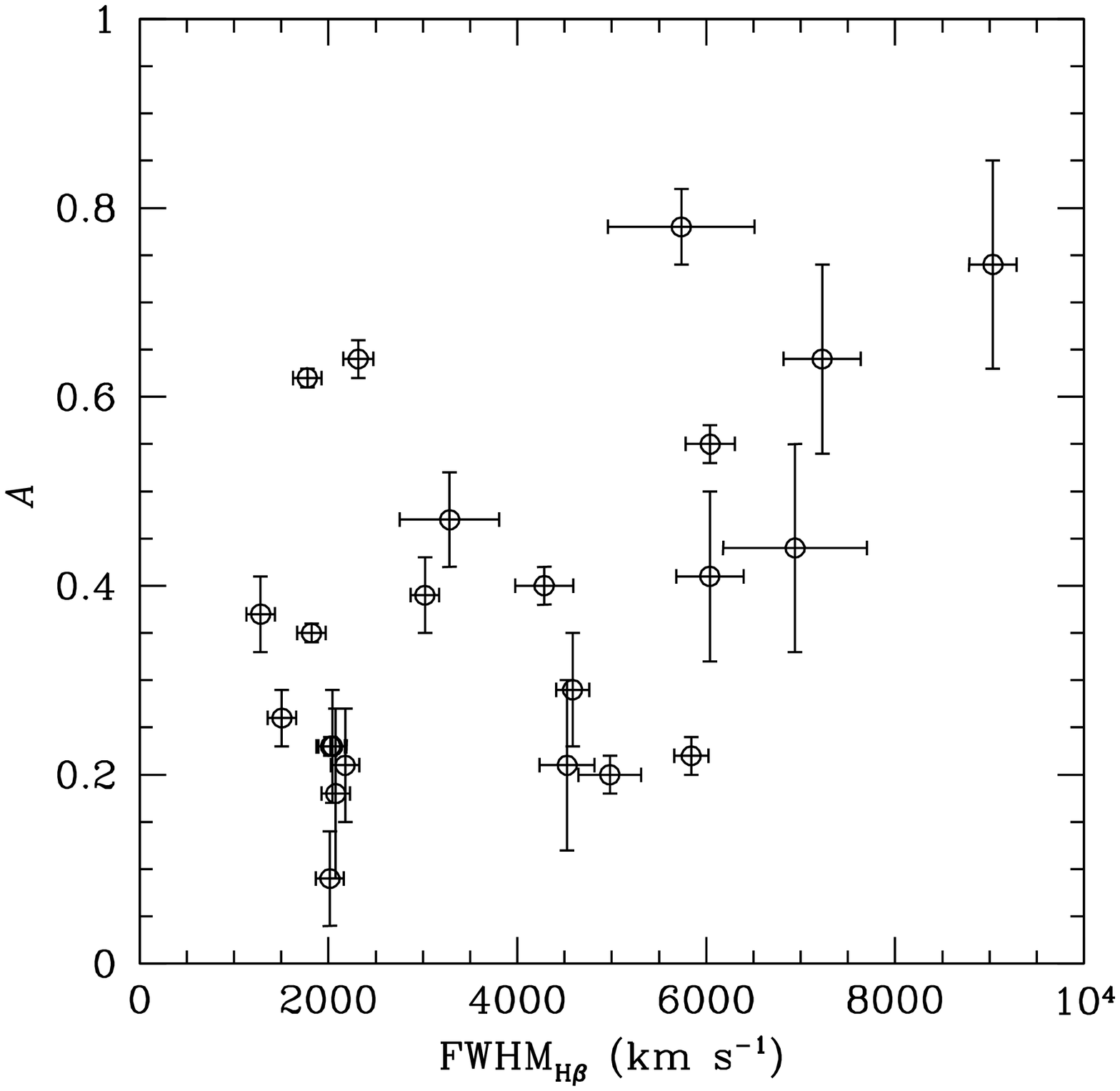}
\includegraphics[width=0.48\textwidth]{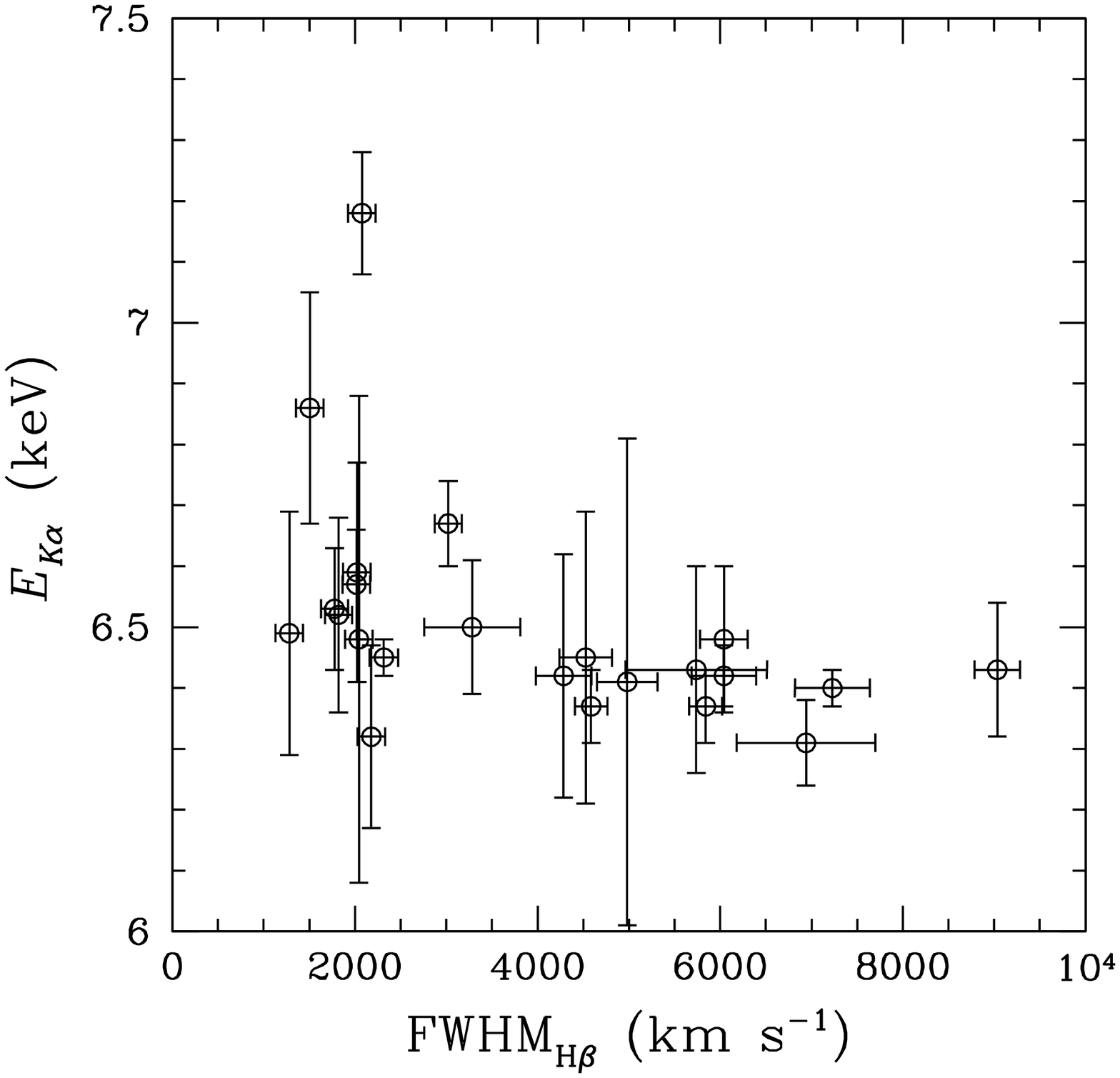}
\caption{{\bf Left panel:} Boltzmann plot slope, determined from the Balmer line intensity ratios, as a function of H$\beta$\ line profile width. {\bf Right panel:} Energy of the X-ray Fe emission as a function of FWHM$_{\rm H\beta}$. The detection of higher ionization in the BLR plasma of narrow line emitting sources is associated to the observation of an averagely higher ionization of Fe in objects with FWHM$_{\rm H\beta} \leq 4000\, {\rm km\, s^{-1}}$.}
\end{figure}
{\it Reverberation Mapping} (RM) investigations demonstrated that the size of the BLR is related with the luminosity of the continuum and that, to some extent, the width of the H$\beta$\ emission line acts as a proxy of the gas velocity distribution. In our previous works \citep[see][]{LaMura07, LaMura09a}, we investigated how the emission line profiles of Type 1 AGNs could be related to the structure of the BLR and how the accretion of material into the gravitational field of sources with different central masses led to various conditions of ionization in the BLR plasma. In particular, it was observed that a higher degree of ionization was found in sources with narrow line profiles, where black holes with lower masses and higher accretion rates were also expected. In this work, we tested those conclusions and the related assumptions on BLR structural models, based on the RM results, by comparing different indicators of matter ionization, inferred from optical and X-ray observations, with the observed emission line profiles. By applying the BP to the Balmer line series, we observe a trend of increasing ionization in the BLR plasma of objects with narrower emission line profiles, corresponding to a correlation coefficient $R = 0.572$ between $A$ and FWHM$_{\rm H\beta}$ and a null hypothesis probability of $P_0 = 0.005$. In other words, when $A$ is decreasing, as it is shown in the left panel of Fig.~3, the normalized intensities $I_n$ of lines from high levels are increasing, as we should expect when less neutral atoms are present along the path of the photons emitted in the considered transitions. The indications of an increasing degree of ionization in the BLR plasma is associated with the detection of averagely higher Fe K$\alpha$\ rest frame energies in objects with FWHM$_{\rm H\beta} \leq 4000\, {\rm km\, s^{-1}}$, suggestive of a higher ionization of Fe, too. This trend, however, is much weaker, with $R = -0.197$ and $P_0 = 0.381$, values which are definitely unacceptable to claim any correlation. This is in part expected, since objects with FWHM$_{\rm H\beta} > 4000\, {\rm km\, s^{-1}}$ simply show a Fe~K$\alpha$\ emission from \ion{Fe}{I} and then, only when FWHM$_{\rm H\beta} \leq 4000\, {\rm km\, s}^{-1}$, the ionization of Fe increases with the energy of the Fe~K$\alpha$. The trend is illustrated in the right panel of Fig.~3, though the small number of available objects does not yet allow to give a statistically conclusive result.

\section{Conclusions}
Clearly much more work is still required in order to assess the existence of an actual relationship between optical line profiles and ionization degree. The problem is further complicated by the fact that we are likely comparing spectral features that arise in different regions of the AGN and, possibly, at different epochs \citep{Shu10}. What, however, looks interesting is the comparison of the ionization properties that we measure by means of optical and X-ray data. While the BLR plasma, that is mainly interacting with the UV / soft-X ray parts of the continuum, shows hints of increasing ionization in narrow line emitters, suggestive of hotter accretion flows around relatively low mass black holes, the detection of highly ionized Fe in regions that are mainly affected by more energetic X-ray photons, points towards averagely higher accretion rates, for sources with FWHM$_{\rm H\beta} \leq 4000\, {\rm km\, s^{-1}}$.

This observation supports the application of structural models, calibrated on the basis of RM results, to the interpretation of physical processes in AGNs and it also gives us useful information concerning the BLR plasma physical conditions in different objects. A more detailed treatment of the problem would certainly benefit from the analysis of a larger sample of objects with broad line profiles characterized by FWHM$_{\rm H\beta} \leq 4000\, {\rm km\, s}^{-1}$, but also from the study of the relationships between X-ray and optical data, through a proper monitoring of the spectral parameters and a correct account of the non simultaneous nature of the observations. In conclusion, our analysis of broad band data, in the absence of systematic effects, such as intrinsic absorptions and extreme geometrical corrections, supports the idea that objects hosting central sources with lower masses tend to be powered by higher accretion rates.




\end{document}